\documentclass[useAMS,usenatbib]{mn2e}

\input epsf

\usepackage{graphicx}
\usepackage{txfonts}
\usepackage{epsfig}

\title[Definitive Sun-as-a-star p-mode frequencies]{Definitive Sun-as-a-star p-mode frequencies: 23 years of BiSON observations}
\author[A-M. Broomhall et. al.]
{A-M. Broomhall, W. J. Chaplin$^1$\thanks{E-mail:
wjc@bison.ph.bham.ac.uk}, G. R. Davies$^1$, Y. Elsworth$^1$, S. T.
Fletcher$^2$, \newauthor S. J. Hale$^1$, B. Miller$^1$ and R. New$^2$\\
$^1$ School of Physics and Astronomy, University of Birmingham,
Edgbaston, Birmingham, B15 2TT, UK\\
$^2$ Materials Engineering Research Institute, Faculty of Arts,
Computing, Engineering and Science, Sheffield Hallam University,\\
Howard Street, Sheffield, S1 1WB, UK }
\begin{document}

\maketitle

\begin{abstract}
We present a list of ``best possible" estimates of low-degree p-mode
frequencies, from 8640 days of observations made by the Birmingham
Solar-Oscillations Network (BiSON). This is the longest stretch of
helioseismic data ever used for this purpose, giving exquisite
precision in the estimated frequencies.  Every effort has been made
in the analysis to ensure that the frequency estimates are also
accurate. In addition to presenting the raw best-fitting frequencies
from our ``peak bagging" analysis, we also provide tables of
corrected frequencies pertinent to the quiet-Sun and an intermediate
level of solar activity.
\end{abstract}


\section{Introduction}\label{sec_intro}

It has been known for over forty years that the surface of the Sun
oscillates with a period of around 5 minutes
\citep[see][]{Leighton1962,Evans1962}. Originally thought to be
local in nature, observations of the 5-minute oscillations using
unimaged sunlight (i.e., Sun-as-a-star observations) proved that the
oscillations were actually a global phenomenon \citep[see
e.g.,][]{Claverie1979}. The oscillations are the detectable
manifestation of acoustic waves that are generated by turbulent
motion in the solar convection zone. Waves of certain frequencies
interfere constructively leading to a large number of resonant p
modes, so called due to the gradient of pressure being the dominant
restoring force \citep[see][]{Ulrich1970,Leibacher1971}.

In Helioseismology, estimates of the frequencies of solar p modes
are used as inputs to the inverse problem of trying to better
constrain the parameters of solar models \citep[see
e.g.,][]{C-Dalsgaard2002}. Hence it is important to determine the
frequency estimates to high precision and high accuracy.

The different resonant modes are characterised by three integers:
the radial order, $n$, angular degree, $\ell$, and azimuthal order,
$m$. Low-$\ell$ modes travel deep into the solar interior, while
higher-degree modes are constrained within the outer regions. Hence,
by determining frequencies of a full range of modes it is possible
to use inversions to determine the structure of the solar interior
as a function of depth. However, the constraints on these inferences
become poorer the deeper one goes into the interior as there are
fewer and fewer modes available which penetrate to successive
depths. Hence when dealing with low-$\ell$ modes, which are observed
most effectively by global (i.e., Sun-as-a-star) observations,
obtaining precise frequency estimates is particularly desirable.

The precision with which any p-mode frequency can be determined is
directly related to the length of observations one makes. The
Birmingham Solar-Oscillations Network (BiSON)
\citep[see][]{Chaplin1996} has collected more than 30 years of
observations of global helioseismology data and is therefore in a
position to provide estimates of low-$\ell$ p modes to unprecedented
levels of precision. In Section~\ref{sec_data} we give details of
the BiSON network and explain how the BiSON time series are
generated. We also give the parameters of the particular data set
used in obtaining the p-mode frequencies given in this paper.

The most common method of determining the frequencies, which is used
here, is to perform a Fourier transform of the time series in order
to generate the frequency power spectrum. The modes are then
parameterized by fitting an asymmetric Lorentzian model to the
peaks, the parameters being frequency, height, width, rotational
splitting and fractional asymmetry. This fitting is colloquially
referred to as ``peak bagging" analysis.

In order to give accurate estimates of these parameters one needs a
model that matches the underlying limit spectrum of the mode peaks
as closely as possible. Here we have employed the ``pseudo-global"
fitting model of \citet{Fletcher2008,Fletcher2009a,Fletcher2009b},
which we describe briefly in Section~\ref{sec_pg_technique}.

The one drawback of using very long time series in order to
determine highly precise frequency estimates is the presence of the
solar cycle. During one activity cycle of the Sun, the p-mode
frequencies change by up to 1~$\mu$Hz (many times the precision we
are attempting to report the frequency estimates to). However, it is
possible to track the change in frequency over time and therefore
the solar cycle effect can be accounted and corrected for. We
present raw fitted frequencies and frequencies corrected for the
solar cycle in Section~\ref{sec_freq} (for the cases of the
quiet-Sun and intermediate levels of solar activity).

\section{The BiSON Data}\label{sec_data}

The BiSON network is comprised of six semi- and fully-automated
solar observing stations that are dedicated to collection of
low-degree (Sun-as-a-star) helioseismic data. The stations are
situated at various sites around the world in order to provide as
continuous observations as possible. The six sites are at: Mount
Wilson in California, Las Campanas in Chile,
Iza$\tilde{\textrm{n}}$a in Tenerife, Sutherland in South Africa,
Carnarvon in Western Australia and Narrabri in New South Wales.

At each of the six stations a resonance scattering spectrometer
(RSS) \citep[see][]{Brookes1978} is used to measure the Doppler
velocity shift of the 770-nm D1 potassium absorption line. This is
done by passing the incident solar light through a cell containing a
vapour of potassium atoms at a temperature of about
$100^\textrm{\footnotesize{o}}$C. Photons of the appropriate energy
are resonantly scattered by atoms in the cell. As the light from
this process is emitted isotropically it is possible, by placing
detectors at right angles to the incident beam, to record only those
photons which have undergone resonance scattering. The absorption
cross section of atoms in the vapour cell is much narrower than that
of the solar Fraunhofer line because the temperature is lower and
there is no rotational broadening. Therefore the intensity of the
recorded light will be proportional to the intensity of light
emitted from a narrow band of the solar absorption line.

The intrinsic sensitivity of the RSS is improved by moving the
passbands out onto the wings of the Fraunhofer line where the slope
is greatest and hence any given line shift will give a greater
change in measured intensity.  The vapour cell is placed in a
longitudinal magnetic field, which causes Zeeman splitting to occur.
The single line is thus split into a multiplet with a separation
dependent on the field strength. The splitting alters the atomic
state in such a way that atoms will interact with circularly
polarized light. The blue-shifted transition is sensitive to one
hand of circular polarization whilst the red-shifted transition is
sensitive to the other.

By using a suitable combination of a linear polariser and a
quarter-wave plate the incident light can be circularly polarised
and hence, by switching quickly between one-handedness and the
other, it is possible to measure the light intensity in the blue
wing, $I_b$ and red wing, $I_r$, almost simultaneously. From these
measurements a ratio, $R$, is formed which gives a near-linear proxy
for the velocity shift of the solar line:
\begin{equation}
R = \frac{I_b - I_r}{I_b + I_r} \label{eq1}
\end{equation}

Finally, to obtain velocity measurements of the solar oscillations,
a third-order polynomial function of the station-Sun line-of sight
velocity is fitted to $R$. The oscillations signal is then recovered
by subtracting $R$ from the polynomial function and calibrated using
the fitted gradient of $R$ versus station velocity.

These measurements of the solar oscillations are collected at each
station and combined to form a continuous time series with a duty
cycle of around 86\% (for 2007 and 2008). Iza$\tilde{\textrm{n}}$a
has been collecting data since 1976, hence there are over 30 years
of BiSON data to call upon. However, with only one station
collecting data and at only certain times of the year the fill for
these early observations is only of the order of 10\%. During the
mid 1980's and early 1990's the remaining BiSON station's were
gradually brought online and the window function improved to around
80\% and above.

In this paper we have chosen to work with an 8640-day set of 40-s
cadence BiSON data running from 4th April 1985 until 16th December
2008. The start date corresponds to a time just after the third
BiSON station came into use in Carnarvon, Western Australia. This
means the data set starts with a fill of around 20-30\% over the
first few years which steadily improves to around 80\% by 1994 once
all six stations were full operational.

\section{The ``pseudo-global" technique}\label{sec_pg_technique}

The best way of determining estimates of the p-mode frequencies is
to take the Fourier transform of the time series in order to get the
frequency power spectrum. For Sun-as-star observations the power
spectrum has a very distinct pattern.  A series of sharp peaks is
seen that coincide with the various p-mode frequencies, with the
largest peaks signifying the strongest modes at around 3000$\mu$Hz
(i.e., a period of around 5 minutes).

On closer inspection the peaks are seen to be grouped in pairs.
These pairs consist alternately of $\ell$~=~0 and 2 modes and
$\ell$~=~1 and 3 modes. In power spectra of long BiSON time series
it is also possible to make out the strongest $\ell$~=~4 modes,
which tend to have frequencies a little lower than the $\ell$~=~2
modes (thus the $\ell$~=~0/2 pair could be considered an
$\ell$~=~0/2/4 triplet). Even closer inspection reveals that that
all but the $\ell$~=~0 modes are rotationally split. However, even
with a very high resolution spectrum this splitting cannot be seen
at high frequencies since modes in this regime have short lifetimes
and consequently their peaks have large widths in the power
spectrum.

To determine the frequency estimates an asymmetric Lorentzian model
is fitted to the various peaks in the frequency power spectrum. The
traditional way of doing this has been to divide the frequency power
spectrum into a series of fitting regions and fit the modes in pairs
(i.e., $\ell$~=~0 and 2 modes together and $\ell$~=~1 and 3
together) \citep[see e.g.,][]{Chaplin1999}. This enables one to
determine how the mode parameters depend on both frequency (overtone
number) and angular degree without the need to fit the entire
spectrum simultaneously.

However, recent work has shown that this method can result in
systematic bias in the returned frequencies
\citep{Jimenez2008,Fletcher2009a}. The reason for this is that the
model only accounts for modes whose central frequencies are within
the fitting region. Therefore, power from modes whose central
frequencies lie outside this region will not be accounted for. This
imperfect match between the fitting model and the underlying profile
can lead to bias in the mode parameters.

A global fitting approach where one fits the entire spectrum
simultaneously would overcome this problem. However it would mean
using a fitting model with many hundreds of parameters which, when
fitting a frequency power spectrum made up of more than 18 million
points, leads to prohibitively long computing times. Also, such a
complicated model may lead to an increased possibility of premature
convergence.

Therefore, in order to eliminate the bias the ``pseudo-global"
method of \cite{Fletcher2009a} was employed. This approach works by
first setting up a model for the full spectrum (FSM) using the
parameter estimates returned from the traditional ``pair-by-pair"
method. Then, as with the pair-by-pair approach, the frequency power
spectrum is divided into a series of fitting regions. Each region is
then fitted using the FSM, but only the parameters of the modes
within each particular fitting region are allowed to vary. In this
way we limit the number of parameters being varied at any one time,
while still including information in the model from all the modes. A
much more detailed account of this fitting method can be found in
\cite{Fletcher2009a}.

Two further improvements were made in the pseudo-global model used
here compared with the techniques used when determining previously
published BiSON frequencies. The first is the inclusion of
$\ell$~=~4 modes in the model. Secondly we take account of the gaps
in the BiSON data by convolving the fitting model with the spectral
window, as opposed to simply fitting for the prominent diurnal
sidebands \cite[see][]{Fletcher2009b}.

When employing the pseudo-global method the fitting regions were
130$\mu$Hz wide and centered at the midpoint of the $\ell$~=~0/2 and
$\ell$~=~1/3 pairs. The $\ell$~=~4 modes were also included in the
model when these modes were visible. Within each fitting region the
following equation is used to fit the power spectral density, $P$,
as a function of frequency, $\nu$:
\begin{equation}
P(\nu)=\left( P'(\nu)+n(\nu)+\sum_{\ell m} \frac{f_{\ell
m}A_l(1+2bx_{\ell m})}{1+x_{\ell m}^2} \right) * W,
\end{equation}
where,
\begin{equation}
x_{\ell m}=\frac{2[\nu-(\nu0_\ell+m\delta\nu_\ell)]}{\Gamma_\ell}.
\end{equation}
$W$ is the spectral window function and the asterisk denotes a
convolution. $P'$ is the power from the remainder of the power
spectrum outside of the fitting region. $A_l$ denotes the mode
heights and $f_{\ell m}$ the relative $m$ component fractional
height ratios (assumed to take fixed values), $b$ is the fractional
asymmetry of the modes (fixed across the fitting region),
$\nu0_{\ell}$ are the mode frequencies, $\delta\nu_l$ are the
rotational splittings and $\Gamma_\ell$ are the mode widths (fixed
across $m$ components). Finally, $n$ is the background term given by
the \citet{Harvey1985} model to account for the solar granulation
with a constant term, $\beta$ to account for instrumental noise.
\begin{equation}
n(\nu) = \frac{2\sigma^2\tau}{1+(2\pi\nu\tau)^2} + \beta,
\label{Harvey}
\end{equation}
where $\tau$ and $\sigma$ represent the time constant and standard
deviation respectively of the granulation signal. Within the fitting
region only the $\beta$ term is allowed to vary. The values of
$\sigma$ and $\tau$ were fixed beforehand, by fitting the Harvey
model to the low and high-frequency regions of the frequency power
spectrum (i.e., outside the region of the modes).

Once all the fitting regions had been covered, an improved set of
parameter values (compared to those returned from the traditional
fitting method) will have been determined for the entire spectrum.
These new values can then be used to set up another model for the
entire spectrum and the whole pseudo-global process can be run
again. In general the process only needs to be repeated one or two
times before no further significant improvements are seen in the
mode parameters between successive iterations.

Differences between the frequency estimates from the pseudo-global
technique and the traditional approach are small. However, because
of the high precision of the BiSON frequencies the differences are
significant and also systematic. The frequencies from the
pseudo-global technique are between 0.02~-~0.04~$\mu$Hz
(1~-~2~sigma) lower than those determined from a traditional
approach over the region 2200~-~3500~$\mu$Hz.

\section{A definitive list of BiSON frequencies}\label{sec_freq}

The raw, best-fitting frequencies from our pseudo-global
peak-bagging analysis are listed in Table~1. We have augmented this
list at very low frequencies with previously published BiSON
frequencies that come from work where we deliberately targeted the
low-frequency regime \citep{Chaplin2002,Chaplin2007c}. We have
included one additional frequency to those previously published, the
$n$~=~7, $\ell$~=~1 at 1185.592~$\mu$Hz, however only the $m$~=~-1
component was detected. To estimate the frequency centroid we
therefore added to the frequency of the observed component the mean
rotational frequency splitting of nearby non-radial modes.

Tables of frequencies have to be used with care, because p-mode
frequencies vary with the level of solar activity. We therefore
provide sufficient information to enable our frequencies to be
corrected to an activity level of choice. This will help comparisons
and combinations with observations made by other instruments, which
do not span the 8640-day period covered here. We present two tables
of frequencies, which were produced by applying suitable correction
procedures to the raw frequencies listed in Table~1. We use the
10.7-cm radio flux \citep{Tapping1990} as our adopted proxy of the
global level of solar activity. We have not corrected the $\ell$~=~4
frequencies, for the reasons described near the end of this section.
Corrections have not been applied to the very low-frequency modes
($<$1450~$\mu$Hz), as these show very little cycle variation and the
corrections would be insignificant at the level of precision of the
data.

\begin{table*}
\centering \caption{Raw, best-fitting frequencies (in $\mu$Hz) from
our pseudo-global peak-bagging analysis.}
\begin{tabular}{cccccc}
\hline n & $\ell$=0 & $\ell$=1 & $\ell$=2 & $\ell$=3 & $\ell$=4 \\
\hline
 6 &     972.613 $\pm$  0.002 &                          &                           &                           &                          \\
 7 &                          &    1185.592 $\pm$  0.004 &                           &                           &                          \\
 8 &    1263.162 $\pm$  0.012 &    1329.629 $\pm$  0.004 &     1394.680 $\pm$  0.011 &                           &                          \\
 9 &    1407.481 $\pm$  0.006 &    1472.836 $\pm$  0.005 &     1535.861 $\pm$  0.008 &     1591.575 $\pm$  0.014 &                          \\
10 &    1548.336 $\pm$  0.007 &    1612.723 $\pm$  0.007 &     1674.540 $\pm$  0.008 &     1729.092 $\pm$  0.016 &                          \\
11 &    1686.601 $\pm$  0.011 &    1749.285 $\pm$  0.007 &     1810.314 $\pm$  0.009 &     1865.307 $\pm$  0.019 &                          \\
12 &    1822.203 $\pm$  0.011 &    1885.091 $\pm$  0.009 &     1945.816 $\pm$  0.013 &     2001.265 $\pm$  0.017 &                          \\
13 &    1957.431 $\pm$  0.012 &    2020.810 $\pm$  0.010 &     2082.131 $\pm$  0.015 &     2137.821 $\pm$  0.019 &                          \\
14 &    2093.535 $\pm$  0.013 &    2156.812 $\pm$  0.014 &     2217.698 $\pm$  0.018 &     2273.563 $\pm$  0.026 &    2324.163 $\pm$  0.051 \\
15 &    2228.774 $\pm$  0.014 &    2292.032 $\pm$  0.015 &     2352.280 $\pm$  0.017 &     2407.707 $\pm$  0.025 &    2458.716 $\pm$  0.076 \\
16 &    2362.823 $\pm$  0.015 &    2425.650 $\pm$  0.014 &     2485.948 $\pm$  0.019 &     2541.754 $\pm$  0.023 &    2593.197 $\pm$  0.053 \\
17 &    2496.226 $\pm$  0.016 &    2559.235 $\pm$  0.015 &     2619.761 $\pm$  0.018 &     2676.252 $\pm$  0.022 &    2728.502 $\pm$  0.038 \\
18 &    2629.724 $\pm$  0.014 &    2693.437 $\pm$  0.013 &     2754.551 $\pm$  0.016 &     2811.440 $\pm$  0.020 &    2864.349 $\pm$  0.033 \\
19 &    2764.211 $\pm$  0.014 &    2828.258 $\pm$  0.013 &     2889.708 $\pm$  0.016 &     2947.075 $\pm$  0.018 &    3000.227 $\pm$  0.034 \\
20 &    2899.101 $\pm$  0.012 &    2963.447 $\pm$  0.012 &     3024.839 $\pm$  0.016 &     3082.439 $\pm$  0.022 &    3136.105 $\pm$  0.032 \\
21 &    3033.845 $\pm$  0.012 &    3098.287 $\pm$  0.013 &     3159.997 $\pm$  0.017 &     3217.881 $\pm$  0.024 &    3271.864 $\pm$  0.048 \\
22 &    3168.726 $\pm$  0.014 &    3233.321 $\pm$  0.016 &     3295.275 $\pm$  0.021 &     3353.580 $\pm$  0.037 &    3408.325 $\pm$  0.069 \\
23 &    3303.652 $\pm$  0.019 &    3368.689 $\pm$  0.021 &     3431.006 $\pm$  0.030 &     3489.699 $\pm$  0.049 &                          \\
24 &    3439.152 $\pm$  0.027 &    3504.397 $\pm$  0.028 &     3567.248 $\pm$  0.041 &     3626.408 $\pm$  0.073 &                          \\
25 &    3575.083 $\pm$  0.046 &    3640.791 $\pm$  0.036 &     3703.388 $\pm$  0.065 &     3762.917 $\pm$  0.102 &                          \\
26 &    3710.938 $\pm$  0.086 &    3777.436 $\pm$  0.050 &     3840.118 $\pm$  0.143 &     3900.029 $\pm$  0.155 &                          \\
27 &    3847.250 $\pm$  0.177 &    3913.995 $\pm$  0.066 &     3977.388 $\pm$  0.297 &                           &                          \\
28 &    3984.507 $\pm$  0.323 &                          &                           &                           &                          \\
\hline
\end{tabular}
\end{table*}

The list of frequencies in Table~2 is pertinent to an intermediate
level of solar activity, which corresponds to a mean level of the
10.7-cm radio flux of $118 \times 10^{-22}\,\rm W\,m^{-2}\,Hz^{-1}$.
The list in Table~3 is pertinent to the canonical ``quiet-Sun''
level of the radio flux, which, from historical observations of the
index, is set at $64 \times 10^{-22}\,\rm W\,m^{-2}\,Hz^{-1}$
\citep[again, see][]{Tapping1990}. To obtain low-degree frequencies
pertinent to a level of activity higher than quiet-Sun, a linear
interpolation can be performed between the values in Table~2 and
Table~3 (or a linear extrapolation may be carried out if the chosen
activity level is higher than $118 \times 10^{-22}\,\rm
W\,m^{-2}\,Hz^{-1}$). Both frequency tables have been produced so
that the frequencies correspond to the \emph{frequency centroids} of
the low-degree modes. The frequency centroids carry information on
the spherically symmetric component of the internal structure.

\begin{table*}
\centering \caption{Corrected frequencies (in $\mu$Hz), pertinent to
an intermediate level of solar activity (see text).}
\begin{tabular}{ccccc}
\hline n & $\ell$=0 & $\ell$=1 & $\ell$=2 & $\ell$=3 \\
\hline
 6 &     972.613 $\pm$  0.002 &                           &                           &                           \\
 7 &                          &     1185.592 $\pm$  0.004 &                           &                           \\
 8 &    1263.162 $\pm$  0.012 &     1329.629 $\pm$  0.004 &     1394.680 $\pm$  0.011 &                           \\
 9 &    1407.481 $\pm$  0.006 &     1472.836 $\pm$  0.007 &     1535.862 $\pm$  0.007 &     1591.575 $\pm$  0.014 \\
10 &    1548.335 $\pm$  0.007 &     1612.723 $\pm$  0.007 &     1674.540 $\pm$  0.008 &     1729.092 $\pm$  0.016 \\
11 &    1686.601 $\pm$  0.011 &     1749.285 $\pm$  0.007 &     1810.314 $\pm$  0.009 &     1865.306 $\pm$  0.019 \\
12 &    1822.203 $\pm$  0.011 &     1885.092 $\pm$  0.009 &     1945.815 $\pm$  0.013 &     2001.264 $\pm$  0.017 \\
13 &    1957.433 $\pm$  0.012 &     2020.811 $\pm$  0.010 &     2082.130 $\pm$  0.015 &     2137.820 $\pm$  0.019 \\
14 &    2093.538 $\pm$  0.013 &     2156.813 $\pm$  0.014 &     2217.697 $\pm$  0.018 &     2273.563 $\pm$  0.026 \\
15 &    2228.779 $\pm$  0.014 &     2292.035 $\pm$  0.015 &     2352.280 $\pm$  0.017 &     2407.709 $\pm$  0.025 \\
16 &    2362.831 $\pm$  0.015 &     2425.655 $\pm$  0.014 &     2485.950 $\pm$  0.019 &     2541.758 $\pm$  0.023 \\
17 &    2496.237 $\pm$  0.016 &     2559.242 $\pm$  0.015 &     2619.766 $\pm$  0.018 &     2676.259 $\pm$  0.022 \\
18 &    2629.741 $\pm$  0.014 &     2693.450 $\pm$  0.013 &     2754.561 $\pm$  0.016 &     2811.455 $\pm$  0.020 \\
19 &    2764.233 $\pm$  0.014 &     2828.279 $\pm$  0.013 &     2889.726 $\pm$  0.016 &     2947.101 $\pm$  0.018 \\
20 &    2899.133 $\pm$  0.012 &     2963.478 $\pm$  0.012 &     3024.867 $\pm$  0.016 &     3082.471 $\pm$  0.022 \\
21 &    3033.886 $\pm$  0.012 &     3098.327 $\pm$  0.013 &     3160.028 $\pm$  0.017 &     3217.916 $\pm$  0.024 \\
22 &    3168.773 $\pm$  0.014 &     3233.358 $\pm$  0.016 &     3295.301 $\pm$  0.021 &     3353.602 $\pm$  0.037 \\
23 &    3303.698 $\pm$  0.019 &     3368.721 $\pm$  0.021 &     3431.019 $\pm$  0.031 &     3489.712 $\pm$  0.049 \\
24 &    3439.195 $\pm$  0.027 &     3504.417 $\pm$  0.028 &     3567.255 $\pm$  0.041 &     3626.420 $\pm$  0.073 \\
25 &    3575.121 $\pm$  0.046 &     3640.799 $\pm$  0.036 &     3703.386 $\pm$  0.065 &     3762.911 $\pm$  0.102 \\
26 &    3710.972 $\pm$  0.086 &     3777.430 $\pm$  0.050 &     3840.096 $\pm$  0.143 &     3900.013 $\pm$  0.155 \\
27 &    3847.277 $\pm$  0.177 &     3913.974 $\pm$  0.066 &     3977.350 $\pm$  0.297 &                           \\
28 &    3984.527 $\pm$  0.323 &                           &                           &                           \\
\hline
\end{tabular}
\end{table*}

\begin{table*}
\centering \caption{Corrected frequencies (in $\mu$Hz), pertinent to
the canonical ``quiet Sun" (see text).}
\begin{tabular}{ccccc}
\hline n & $\ell$=0 & $\ell$=1 & $\ell$=2 & $\ell$=3 \\
\hline
 6 &     972.613 $\pm$  0.002 &                           &                           &                           \\
 7 &                          &     1185.592 $\pm$  0.004 &                           &                           \\
 8 &    1263.162 $\pm$  0.012 &     1329.629 $\pm$  0.004 &     1394.680 $\pm$  0.011 &                           \\
 9 &    1407.481 $\pm$  0.012 &     1472.835 $\pm$  0.007 &     1535.859 $\pm$  0.011 &                           \\
10 &    1548.333 $\pm$  0.007 &     1612.720 $\pm$  0.007 &     1674.534 $\pm$  0.008 &     1729.085 $\pm$  0.016 \\
11 &    1686.597 $\pm$  0.011 &     1749.279 $\pm$  0.007 &     1810.304 $\pm$  0.009 &     1865.293 $\pm$  0.019 \\
12 &    1822.196 $\pm$  0.011 &     1885.081 $\pm$  0.009 &     1945.797 $\pm$  0.013 &     2001.243 $\pm$  0.017 \\
13 &    1957.421 $\pm$  0.012 &     2020.793 $\pm$  0.010 &     2082.102 $\pm$  0.015 &     2137.788 $\pm$  0.019 \\
14 &    2093.518 $\pm$  0.013 &     2156.784 $\pm$  0.014 &     2217.656 $\pm$  0.018 &     2273.516 $\pm$  0.026 \\
15 &    2228.749 $\pm$  0.014 &     2291.993 $\pm$  0.015 &     2352.222 $\pm$  0.017 &     2407.643 $\pm$  0.025 \\
16 &    2362.788 $\pm$  0.016 &     2425.595 $\pm$  0.015 &     2485.873 $\pm$  0.019 &     2541.671 $\pm$  0.023 \\
17 &    2496.180 $\pm$  0.017 &     2559.162 $\pm$  0.015 &     2619.668 $\pm$  0.018 &     2676.149 $\pm$  0.023 \\
18 &    2629.668 $\pm$  0.015 &     2693.347 $\pm$  0.014 &     2754.439 $\pm$  0.017 &     2811.318 $\pm$  0.021 \\
19 &    2764.142 $\pm$  0.015 &     2828.150 $\pm$  0.014 &     2889.578 $\pm$  0.018 &     2946.935 $\pm$  0.021 \\
20 &    2899.022 $\pm$  0.013 &     2963.322 $\pm$  0.014 &     3024.689 $\pm$  0.018 &     3082.273 $\pm$  0.025 \\
21 &    3033.754 $\pm$  0.014 &     3098.140 $\pm$  0.015 &     3159.821 $\pm$  0.020 &     3217.683 $\pm$  0.028 \\
22 &    3168.618 $\pm$  0.017 &     3233.139 $\pm$  0.018 &     3295.063 $\pm$  0.025 &     3353.335 $\pm$  0.041 \\
23 &    3303.520 $\pm$  0.021 &     3368.469 $\pm$  0.023 &     3430.748 $\pm$  0.034 &     3489.409 $\pm$  0.053 \\
24 &    3438.992 $\pm$  0.030 &     3504.129 $\pm$  0.030 &     3566.949 $\pm$  0.044 &     3626.078 $\pm$  0.075 \\
25 &    3574.893 $\pm$  0.048 &     3640.475 $\pm$  0.038 &     3703.044 $\pm$  0.067 &     3762.530 $\pm$  0.105 \\
26 &    3710.717 $\pm$  0.088 &     3777.067 $\pm$  0.052 &     3839.717 $\pm$  0.144 &                           \\
27 &    3846.993 $\pm$  0.177 &     3913.570 $\pm$  0.068 &     3976.930 $\pm$  0.298 &                           \\
28 &    3984.214 $\pm$  0.323 &                           &                           &                           \\
\hline
\end{tabular}
\end{table*}

The Sun-as-a-star data do not always provide estimates of the
centroids by default. That is because only $\ell+1$ of the $2\ell+1$
components in each non-radial mode are seen clearly in the
observations (those with even $\ell+m$).  At times of low solar
activity, components in the non-radial modes are arranged
symmetrically in frequency: then peak-bagging of the Sun-as-a-star
data \emph{does} give the required centroids. However, at times when
there is significant surface activity -- as in our 8640-d time
series -- the components are not arranged symmetrically in
frequency, and so because some components are missing from the
Sun-as-a-star observations we measure a slightly different frequency
to the centroid. It is possible to make a correction for this
effect, as discussed in detail by \citet{Appourchaux2007}. We have
applied such a correction to give the $118 \times 10^{-22}\,\rm
W\,m^{-2}\,Hz^{-1}$ frequencies in Table~2.

The frequencies in Table~3 are corrected to the quiet-Sun level,
using a procedure that by definition corrects to the expected
quiet-Sun \emph{centroid} frequencies. The procedure assumes that
the frequency correction can be parameterized as a linear function
of the chosen activity proxy. The BiSON correction is
self-consistent. First, we divide the long BiSON time series into
shorter pieces in order to quantify the sensitivity of the
frequencies to the changing 10.7-cm flux (by linear regression). We
then use this information to calibrate our correction. Since we
could not obtain robust estimates of the $\ell$~=~4 frequency shifts
using our data, we did not attempt to correct the frequencies of
these modes. There are gaps in the BiSON time series (e.g., due to
inclement weather, and very occasionally instrument problems). The
window function of the BiSON observations was therefore also applied
to the time series of the 10.7-cm radio flux.

Full details of the correction procedures will be published
elsewhere (Broomhall et al., in preparation).

\section*{Acknowledgments}

We would like to thank all those who are, or have been, associated
with BiSON. BiSON is funded by the Science and Technology Facilities
Council (STFC).

\bibliography{fletcher}

\end{document}